# Using HistCite software to identify significant articles in subject searches of the Web of Science. [1]

Enrique Wulff Barreiro
Marine Science Institute of Andalucía (CSIC)
Avda. República Saharaui, s/n
11510-Puerto Real (Cádiz). Spain.
e-mail: enrique.wulff@icman.csic.es

## ABSTRACT[2]

HistCite[TM] is a large-scale computer tool for mapping science. Its power of visualization combines the production of historiographs on the basis of the analysis of co-citations of documents, with the use of bibliometrics specific indicators. The objective of this article is, to present the advantages of the new bibliometrics configuration of HistCite[TM] (2004) when identifying articles. The analysis of the histograms that produces HistCite[TM], in terms of cumulative advantage and aging of the citations. And the comparative study of the results of HistCite[TM], in its indicators of amplitude and recognition. Also is examined its treatment of the sampling problems, by formalizing the question of Kendall.

---

[1] The readers should contact Eugene Garfield <eugene.garfield@thomson.com> and Soren Paris <soren.paris@contractor.thomson.com> if they wish to gain access to the software. They should know that the version of HistCite which has been used for this article is now displaced by a more up to date version. HistCite is a work in progress.

[2] A Spanish informational abstract is available from: http://www.cindoc.csic.es/info/fesabid/27.htm.

## INTRODUCTION

The program by E. Garfield (emeritus president of Thomson ISI), A.I. Pudovkin (biologist at the Institute of Marine Biology, Vladivostok) and V.S. Istomin (formerly at Washington State University, now in Vladivostok), HistCite[TM] is a new tool, that operates from 2002 in the arena of the international scientific edition. On November 18, 2002, the authors presented this software at the 65th annual conference of the American Society for Information Science & Technology (ASIST) along the bibliometrics session (where a work on the psychologists Kurt Lewin was exposed, an analysis of 20 years of the journal Scientometrics was displayed, and an inquiry of the medical informatics market was exposed).

With the study "Towards a model for science indicators" [1] Derek John de Solla Price (1978), the mentor of the scientific humanities [2], started a research program where humanities came to the social sciences. In effect, the origin of the exploration of scientific theories in search of change signals targets the identification of crucial experiments logically "requiring" the reception of new representations (decidibles in terms of continuity or rupture). The coherent use of the bibliographic resources in order to develop serious cases with which to face the truth of a hypothesis is a coherent historical practice (being a classical case the one that refers the expression of Kapitza [3] "The publications had been interrupted!" [3]). One more step takes place in 1964 when Garfield [4] underlines the quantitative capacity that shows the computer science to assemble cases as a viable narrative equivalence by using the book of history of Isaac Asimov "The genetic code".

The objective of HistCite[TM] is to fit the tensions in the theoretical informative bridge, by identifying the intellectual constitution of the collective centers of attention for a period of given time. We understand by "theoretical" the implicit idea of "gap" (synapsis) by contrast to the capillar concept, that opposed Ramón y Cajal to Golgi with regard to the "brain bookeeping economy". Bridging the gap is the tailoring of a common denominator, is serious negotiations between supplier and consumer. By interpreting these last two terms in the sense of the input-output matrices by Wassili Leontief, considered in the triple dimension to read-to-cite-to-write. The management of this denominator as a "truth moment" gives rise to scale economies for the organizations involved [5]. The resulting suggestions become indicators when the passion that inspires those in charge with the process of evaluation concedes the filling of the gap between narration and history  (Henry Small indicates that "… the specialty narrative is not a history of the specialty" [6]).

Strictly speaking it would have to be accepted that about the statistical results only the peaks of the editorial activity will be observed. As Edward Nadel [7] points out even an estimation of tendencies (for pairs of cited articles that are cited together, or cocitations) is doubtful, on the basis of models of regression  with a dependent variable that uses dichotomous data (like the citations, because an article is or not cited). Nevertheless, the absence of singularity that emerges from the matrix of correlation, as a result of the homoscedasticity principle violation for the variance error, is afforded without significant differences, as much when factorial analysis is used (for the extraction of the initial factor) like when the results are obtained by using principal components extraction. If a different information source is consulted (like the opinion of scientific experts in a field), the labelage by means of the allocation of factors is adequate without criticisms, at least to divide the history into periods of time. The list of "received" retrospective participants that therefore is obtained differs, and it looks like, to the indicators of immediate formal activity.

Under the assumptions of biological evolution, the scientific evaluation progresses confronted with the business world [8]. Because it equalizes the market activity with the development of knowledge centers [9]. Nevertheless, due to the economic volatileness environmental criteria (individuals do not stop searching for fresh opportunities) [10], when elaborating numerical simulations of the highly connected society the differences between the majority of the social networks and the networks of scientific collaboration must be observed. In the general networks the capacity of the agents to maintain or enlarge the number of its contacts is severely limited by the constant and unavoidable action of deterioration. Instead of that, the collaboration in scientific networks deals with a kind of network (biological, technological or informational) whose links embody a "one-and-for-all-choice" (where in contrast with the ephemeral life of the authors the papers, once written, do exist for ever), and therefore they are more stable. Nevertheless in order to activate the growth in a subsequent phase it is necessary the intervention of a nucleation event. In this vein the network structure connectivities properties are exploited (small world properties [11], degree distributions of power law, and network transitivity [12]).

The method that analyzes the general dynamics of a network, its connectivity, is the search algorithm. Associated to a universe of publications this method analyzes a citation network to identify a set of articles with a central role in the development of a theory [13]. The interactive attractiveness in the making-up of a network is guaranteed by the effectiveness of the search, that is associated to the homeostatic production of an image.

The visualization of an informative space, associated to an effective search, standardise the processes of elaboration of virtual libraries. This is the sense for the elaboration of scientometrics indicators [14] for the scientific evaluation that is centred in the inference and the reduction of the search permitted by the presentation of the data in an image [15]. Furthermore, the theoretical development of the vision by means of the computer regularizes the fugue point towards are lined the automatized searches (articulated by the Red Queen Principle own to evolutive biology [16][17]).

## BIBLIOMETRICS INDICATORS OF THE HISTCITE[TM] 2004 VERSION

The HistCite[TM] software is a system for the historiography analysis that organizes the bibliographic collections generated by searching in the Science Citation Index of the Web of Science (WOS) or in the SCI-CD-Rom. It permits to follow the evolution of articles, authors, and journals and the graphical representation of the more influentials articles on a subject chronology. The present contribution is based in its 2004 version.

The HistCite[TM] bibliometrics viewer offers several statistics indicators.

GCS, the global citation score based on the ISI Web of Science (WOK) database record.

GCS/t, or the global citation annual score, facilitates the frequency of the annual citation based on the Web of Science reckoned at the moment when the data was downloaded.

TGCS, total score in ISI Web of Science (WOK) for all the publications of an author or source. (It appears at the lists of authors and sources.)

LCS, local citation frequency inside the collection.

LCS/t, or local annual citation score, is the annual occurrences of citations received by an article from other papers inside the local collection that have been downloaded for the study of the specific case.

TLCS, total calculation of the local citation frequency inside the collection. (It appears at the lists of authors and sources.)

NCR, amount of mentioned references. It presents the references included in the bibliography of an article. (It appears at the matrix of citations.)

LCR, mentioned local references. It presents the articles referred that are part of the collection. (It appears at the matrix of citations.)

$LCS_b$, or local citations score in the beginning. It shows the local citation frequency inside the collection (LCS), or local citations score in the beginning. It shows the local citation frequency inside the collection (LCS), only for a period including from the moment when the collection was compiled, until an arbitrary cutoff year in the future.

$LCS_e$, or local citations score in the end. It presents the local citation frequency inside the collection (LCS) only for a period that starts with an arbitrary cutoff year until and including the last year of the time interval for which the collection has been compiled.

## THE HISTCITE[TM] HISTOGRAMS IN VIEW OF THE EFFECTS OF THE AGING AND CUMULATIVE ADVANTAGE OVER THE CITATION NETWORKS AND CO-AUTHORS. WEIBULL DISTRIBUTION.

The idea of history of citation of an article is introduced after the work by Jan Vláchy (1989 Price medal [18]). In order to approximate the measure of citation aging like a major argument for the future of a research program. A history of citations is a temporal citation profile for all the articles published in a year, by a person or on a subject. The histories of citations are different in regard to two typologies: intuitive and formalized. According to the intuitive typology we can distinguish between articles of normal acknowledgment, articles of gradual acknowledgment, innovating articles with late acknowledgment, erroneous articles but with good reception and genial articles. The formalized typology (exponential-exponential) distinguishes between the highly acknowledged articles in the beginning, the articles with basic acknowledgment, the articles with scarce echo, the erroneous articles but received well, and the genial articles. These typologies do follow the statistical interpretation of the transitional probabilities that appears in the empirical distribution of the citations, and that makes different from the normal distribution or Gaussian (points out Dr Price [19]). It is understood that the citation histories are statistically meaningful in terms of cumulative advantage.

Nevertheless, the assimilation of this phenomenon to an specific economic behaviour has been estimated by "minor participants" [20] and its algorithmic approach has been qualified as nonexistent [21].

**Additivity of the levels of reference**

In this contribution the cumulative effects produced by HistCite[TM] has been examined. If it is said that the zero reference level is the universe of articles that an author has read to write a text, his reading domain, the mission of the science historian is to trace all the connections between the read texts, purposely for a research problem, in such a way that he can underline the line forces guiding its elaboration. The advantage of this kind of algebrization is to be able to avoid all the superfluous relationships between the citations, emphasizing a non-accumulative essential structure that reproduces the methodological innovations trajectory in the field. [22]

The author of this communication has presented the informative effects on an epidemiological alert along the 1992 Barcelona Olympic Games (http//isdm.univ-tln.fr/PDF/isdm2/isdm2a12_barreiro.pdf) inside the journal ISDM  of the Université du Sud Toulon-Var. We use the concept of additivity after Shapley [33], ie, by assimilating it to linearity (a magnitude depends on others which are the result of a sum). Thus, if in the zero reference level an author must only refer in his publications to his reading domain, in the next level he will be able to include in his references lists the publications that were grouped in the zero level, plus the material that was referred inside the bibliographies of the zero level. The activity in local network of the authors finds therefore a model that follows the pattern of the analysis of references [24].

HistCite[TM], with 15 (identified by WOK) over 17 of the bibliographic references included into the "Annexe", of the mentioned article, defines a citation matrix like the one presented in the Table I.

**Table I: Citation matrices in the area of the Steptococcus pneumoniae epidemiological alert, along the Barcelona '92.**

Fri Feb 11 18:51:02 2005

Nodes: 15, TLCS: 19, TGCS: 1295, mean TLCS: 1.27, mean TGCS: 86.33
Sorted by year, source, volume, issue, page.

| cited nodes | LCR | NCR | Nodes | LCS | GCS | citing nodes |
|---|---|---|---|---|---|---|
|  | 0 | 8 | 1 1987 PEREZ JL | 3 | 45 | 2 4 9 |
| 1 | 1 | 34 | 2 1987 PALLARES R | 5 | 301 | 4 5 9 10 11 |
|  | 0 | 16 | 3 1988 MENDELMAN PM | 1 | 51 | 6 |
| 1 2 | 2 | 6 | 4 1989 CASAL J | 3 | 16 | 9 10 11 |
| 2 | 1 | 3 | 5 1989 MANRESA F | 1 | 6 | 11 |
| 3 | 1 | 29 | 6 1989 DOWSON CG | 2 | 203 | 7 9 |
| 6 | 1 | 19 | 7 1991 MUNOZ R | 0 | 287 |  |
|  | 0 | 4 | 8 1991 MARTINEZ E | 0 | 4 |  |
| 1 2 4 6 | 4 | 31 | 9 1991 FENOLL A | 2 | 250 | 14 15 |
| 2 4 | 2 | 66 | 10 1992 GARCIALEONI ME | 0 | 95 |  |
| 2 4 5 | 3 | 6 | 11 1992 SANCHEZ C | 0 | 28 |  |
|  | 0 | 3 | 12 1992 MARTINEZ E | 0 | 0 |  |
| 14 | 1 | 6 | 13 1992 BARNETT ED | 0 | 3 |  |
| 9 | 1 | 2 | 14 1992 BARNETT ED | 2 | 6 | 13 15 |
| 9 14 | 2 | 7 | 15 1992 PLASENCIA A | 0 | 0 |  |

Generated by: HistCite(Vlad). Version: 2004.05.14

The article at node Nº9 (see "Nodes" in Table I) includes in its list of references 4 articles (Nº 1,2,4,6) from the local collection (as identified by the indicator LCR (local cited references)). Fenoll and collaborators have read these articles, they are part of his reading domain, of his zero reference level. But, particularly, the article Nº4 is also part of his reading domain, of his zero reference level. But, particularly, the article Nº4 is also part of his one reference level, because it can be admitted that Dr. Fenoll has read the works of Dr. Pérez (art. Nº1) and Dr. Pallares (Nº2) after finding them between those selected by Dr. Casal (art. Nº4) to make reference.

It can also be said that the article by Dr. Pallares (Nº2), that has been cited five times inside the local collection (as can be read in the column LCS – local citation frequency inside the collection, Table I), has the possibility of being included in the reference levels of six authors (he himself and the five having cited him). The resulting citation network identifies in this article (Nº2), generated in the inception year (1987) of this period, the most attractive node for the set of the connections of the network. Its feasibility as reference is maximum, on the basis of the greater number of citations that it receives, as much local (LCS=5), as globally for the complete WOK database (GCS=301).

**Management of the co-authorship**

When studying the co-authorship it can be corroborated how the total number of articles produced per year diminish because two authors when acting together do not produce more than one article. In Table II the authors Casal, Fenoll and Muñoz display the same annual rate of total citations within the collection (TCS/t), and they are used like reference 1,75 times annually by the other 60 authors, along the period 1989-1991 for which data have been reckoned. Nevertheless, Fenoll publishes two papers in co-authorship with Muñoz and Casal, publishing these two last (working together) three each one.

**Table II: Sequence of the 10 first authors in a bibliography on an epidemiological alert published in: http//isdm.univ-tln.fr/PDF/isdm2/isdm2a12_barreiro.pdf (ordered by the citation annual score within the collection, LCS/t).**

Glossary   HistCite Guide

Ranked All-Author list.
Total: 60, TLCS: 19, TGCS: 1295, mean TLCS: 0.32, mean TGCS: 21.58
View: Bibliometric. Sorted by **LCS/t**.

| # | Name | TLCS | TLCS/t | TGCS | TGCS/t | TLCSb | TLCSe | Pubs | TLCR |
|---|------|------|--------|------|--------|-------|-------|------|------|
| 1 | LINARES J | 11 | 2.08 | 555 | 109.92 | 1 | 3 | 4 | 3 |
| 2 | BARNETT ED | 2 | 2.00 | 9 | 9.00 | | 2 | 2 | 2 |
| 3 | KLEIN JO | 2 | 2.00 | 9 | 9.00 | | 2 | 2 | 2 |
| 4 | TEELE DW | 2 | 2.00 | 6 | 6.00 | | 2 | 1 | 1 |
| 5 | CASAL J | 5 | 1.75 | 553 | 272.50 | | 4 | 3 | 7 |
| 6 | FENOLL A | 5 | 1.75 | 266 | 129.00 | | 4 | 2 | 6 |
| 7 | MUNOZ R | 5 | 1.75 | 553 | 272.50 | | 4 | 3 | 7 |
| 8 | DORCA J | 6 | 1.08 | 307 | 51.67 | 0 | 3 | 2 | 2 |
| 9 | PALLARES R | 6 | 1.08 | 307 | 51.67 | 0 | 3 | 2 | 2 |
| 10 | BOURGON CM | 2 | 1.00 | 250 | 125.00 | | 2 | 1 | 4 |

**Cumulative advantage process**

In the graph 1A it is viewed how the aging produces an effect of cumulative advantage in favor of the frequently cited old article Nº2. The same effect results in advantage for the article Nº1; in this way 13% of the initial sources of inspiration is underlined, thanks to the strong thematic induction that results from the aging (although seven positions mediate between the articles Nº1 and 2, from the values of column of GCS (global citation score) at Table I).

The graphical interface allows to fix an annual value like determinant so that a connection by means of a citation visualize an author's paper. In graph 1A the author of the article Nº6 shares ideas (by means of citations) with the author of the article Nº3; when the year 1988 is considered. When requiring a threshold of 55 citations (graph 1B) in the WOK database, the year 1988 no longer it supposes an advantage.

**Graph 1: Transitions between nodes with influence on the year of citation.**
**1A. 0 Citation threshold (Global in WOK);   1B. 55 Citation threshold (Global in WOK)**
**15 Nodes; 19 Links.                                       5 Nodes; 4 Links.**

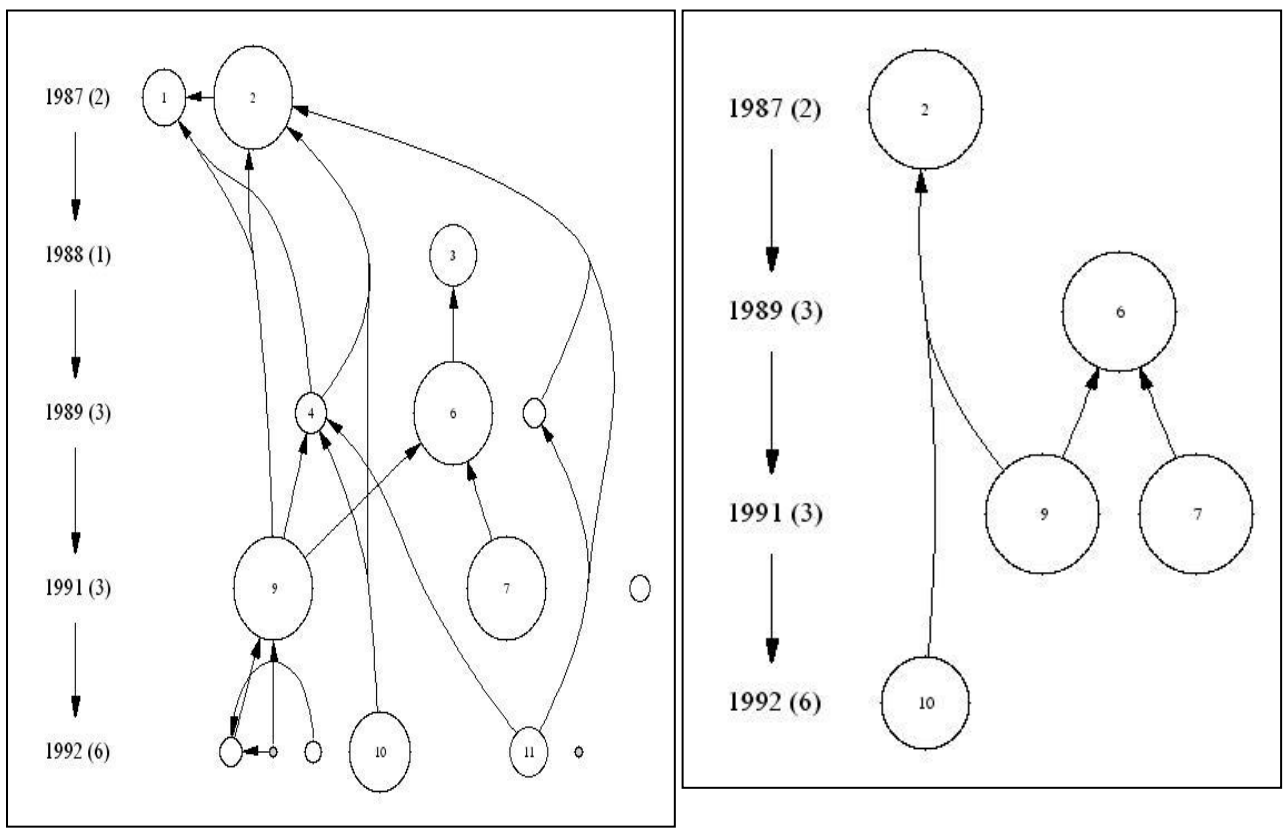

**Egalitarian distribution of acknowledgment, and Weibull distribution**

Jan Vláchy advances, that the genuine relationship between the reference levels and the aging of the citations admit a model that is based on the Weibull distribution (because it turns out suitable to consider temporary data related with human failures, like for instance the no-reception of citations by an article along a period of time). As the obsolescence phenomena is general, and emerges even before the publication of the articles (e.g. because the research continues from a distinct point that the one mentioned by the authors in the citation – the parameter of localization of the citation as a new bibliometrics variable was introduced by Dr. V. Cano [25] [26]), the more egalitarian distribution of the citations would be the result of managing the aging by moderating the cumulative advantage that favors the "old" articles. Different Weibull functions could be generated, according to the aging, in order to determine what should be the value of the parameters in such a way that the more recent articles are favored. This effect suggests that the articles can not only be clustered by subjects but also in a time-dependent way.

With HistCite[TM] the citation threshold can be reduced to zero, resulting in an enhancement of the temporal cluster degree. Note that, in the Graph 1A , in 40% of the local collection, after 1991, it exist a "cluster" specific to the reading behavior "examined closely" (for the three papers published in 1992, at the left side of the graph).

## CARTOGRAPHY OF CASE STUDIES

By incorporating the idea of the cartographic presentation of the bibliographic information, the identification of the scientific authors communities can be considered by dividing them into core, continuants and transients authors. With this aim, one additional methodology for the study of the stability of HistCite[TM], expressed in terms of indicators' production, consists of comparing its benefits with the results, in two case studies published by scientific journals, and in exploiting its performances of graphical modelling.

Interesting coincidences between the cartographies will be considered for its analysis. A comparison will be established between the published charts. So, in the case of the history of the editorial department of a journal in management science and, in the case of the citation histories in cancer research their maps will be compared with the graphs produced by HistCite[TM].

This section finalizes with a normal model for the distribution of the bibliographic references, known as the Kendall question. By consequence, the capacity of the software for the resolution of sampling problems is discussed.

### History of the Publishing Department of the journal 'Management Science'.

The journal of operations research and sciences of the management 'Management Science', of the Northwestern University in Illinois (U.S.A.), discusses in its issue of May 2004 [27], the history of the department of analysis of the decision in 'Management Science', describing its publishing structure from 1970 to the present. In this vein, it reviews articles published, in this journal, on analysis of the decision.

By satisfying the search criteria by source title (Source title = Management science) and subject (topic = decision analysis) an scenario was constructed with 71 articles, indexed by the database ISI Web of Knowledge (WOK) , that elapses along the period 1975-2004. The problem of discernment of the citation cycle that exist inside the bibliography was solved by using HistCite[TM].

A histogram and a frequency table were elaborated. By virtue of the first one (Graph 2), the two indicators of amplitude as used by HistCite[TM] were proven. The local citation score (LCSb) in the beginning of the period is exposed at the axis of ordenates, and in the axis of abcisas the local citation score in the end of the period (LCSe). The table of frequencies (Table III) present the annual acknowledgment indicators of HistCite[TM], LCS/t and GCS/t.

**AMPLITUDE INDICATORS:** The compiled collection is limited temporarily in beginning with $LCS_b$ = 1992 (ie initial computation of citations, inside the local collection, for the period 1975-1992, 17 years (average life of a scientific generation in social sciences) ) and in the end with $LCS_e$ = 1997 (final computation of citations, inside the local collection, for the period 1997-2004, 7 years (registered period to sample estimations of continuity and transcience)). This way, 43.67% of the bibliography it is controlled because to have values different to zero, for both indicators. In such a way, the path longitude in the citation network is maximum. From the synchrony between values of appointment previous and later to fixed dates of initial and final threshold a common coordinate stands out. Thus a common connection through which the network grows is visualized. An implementation of this method described the articles Nº1 and Nº10 as peaks with values distinct to zero for the common coordinates (1,1) and (1,2), of $LCS_b$ and $LCS_e$.

**Graph 2: Management Science : Local citation frequencies at the beginning (1975-1992) and the final of the period (1997-2004).**

(1,0) ≈ art. Nº4,9,13,17,26,34,38,58,62,66
(1,1) ≈ art. Nº 1
(1,2) ≈ art. Nº 10
(2,0) ≈ art. Nº24,36,53,65
(3,0) ≈ art. Nº40,49
(4,0) ≈ art. Nº31

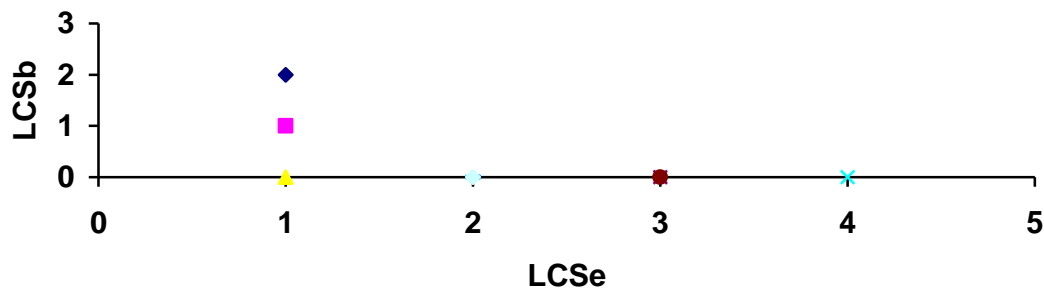

**Pairs of coordenates (LCSb,LCSe) distintct to 0. They model 27% of the bibliografy on 'Management Science'.**

**INDICATORS OF ANNUAL ACKNOWLEDGMENT**: A criterion of comparison apt for a space of common coordinates is also added, by using the annual computation of the citations inside the local collection (LCS/t) and inside the global database WOK (GCS/t). The growth behavior that simulates this representation, is the one to compare the creation of interest generated by the own system of referage (in annual key: LCS/t), with the popularity of the recognition of the activity of the Department of Decision Analysis, when it is applied to the structure of a global community (GCS/t, on an annual base), like the database ISI Web of Knowledge. In the pair thus formed, a solution exists for a behaviour of the uniform growth (where the relevance and the interest personally identified predominates), that limits the cumulative advantage effect, LCS/t; and also it exists a solution for a behaviour of preferential growth, greatly dependent of popularity, GCS/t [28]. Between the two numbers it exist a different qualitative intensity (in terms of the thematic independence). To interpret the model of coincident pairs follows the same abstraction that in the case of the amplitude. The before referred more outstanding articles, are the same than in the estimation according to the amplitude. In effect, the papers N° 1 and N° 10 occupy the first place according to the indicators GCS/t and LCS/t, respectively. An analysis of the thematic category (field 'Subject Category') identify them as theoretical and methodological articles; by extending this consideration to the 22 articles it is verified that its thematic is the research operative and the management science.

**Table III: The observed frequencies in the indicators of annual acknowledgment by HistCite^TM, LCS/t and GCS/t. (N° indicates the node of the article).**

| N° | 12 | 13 | 22 | 9 | 4 | 28 | 62 | 53 | 38 | 58 | 36 | 24 |
|---|---|---|---|---|---|---|---|---|---|---|---|---|
| GCS/t | 0.20 | 0.36 | 0.46 | 0.50 | 0.54 | 0.83 | 0.83 | 0.88 | 0.9 | 1 | 1.09 | 1.17 |
| LCS/t | 0.07 | 0.14 | 0.08 | 0.05 | 0.04 | 0.08 | 0.17 | 0.25 | 0.1 | 0.14 | 0.18 | 0.25 |

| N° | 26 | 31 | 66 | 17 | 40 | 49 | 34 | 10 | 65 | 1 |
|---|---|---|---|---|---|---|---|---|---|---|
| GCS/t | 1.25 | 2.58 | 2.75 | 3 | 3.7 | 3.89 | 3.91 | 4.18 | 4.8 | 7.47 |
| LCS/t | 0.08 | 0.5 | 0.25 | 015 | 0.4 | 0.33 | 0.09 | 0.53 | 0.4 | 0.17 |

The optics adopted here for comparing a published cartography (in 2004) and the results of HistCite^TM, could be synchronous with the potential use of the table 'Missing Links' that the software produces. As WOK frequently introduces articles asynchronously  by the moment of its publication, the data correctness is not absolute. Table IV reports on the pagination of the first reference (that missed in the original list) and re-send towards the correct reference in the second case (because, in this case, the original pagination data were recorded with errors).

**Table IV: 'Missing Links' : Potentially missing citations, and 'variations'**
Potentially missed citations...
2 nodes have citations that may potentially refer to other nodes.

1 | 7 1982 MANAGEMENT SCIENCE 28(3):276-288
NORTH DW; STENGEL DN
*DECISION-ANALYSIS OF PROGRAM CHOICES IN MAGNETIC FUSION ENERGY DEVELOPMENT*

```
SPETZLER CS, 1975, MANAGEMENT SCI, V22 may refer to 1 SPETZLER-CS-
1975-V22-I3-P340-358
```

2 | 51 1997 MANAGEMENT SCIENCE 43(1):1-14
Browne GJ; Curley SP; Benson PG
*Evoking information in probability assessment: Knowledge maps and reasoning-based directed questions*

```
BENSON PG, 1995, MANAGE SCI, V41, P1637 may refer to 43 BENSON-PG-
1995-V41-I10-P1639-1653
```

**The Cartography of the "1988 Dr. Josef Steiner Cancer Research Foundation Award"**

The author of this communication published in 1996 an algorithmic historiography of the research program on ras oncogenes that culminates in the "Nobel Prize" of the cancer research (awarded by the Switzerland research foundation on cancer Josef Steiner) [29]. In the present contribution a discussion is established about if the use of HistCite^TM and of identical bibliographical collection than the employed in the article published in the journal "Llull", can answer for the identification of crucial events and the demographic discrimination of the authors that participate in the research.

The singularity of this comparison comes from the non-frequent convergency of the indicators towards the follow up on a research until the level of the honorific reward. On the other hand, the performances when navigating with HistCite^TM through the information space and in obtaining running details, do interest when determining the specular symmetries between both graphs and the respective internal homogeneities. The identical values that compares the symmetries are the crucial events and the different internal polarities are in correspondence with the demographical classes in which the population is splitted.

The methodology used by the article in 'Llull' is the documental analysis through the use of cocitations. The implemented by HistCite^TM combines the use of indicators (like global and local annual counts) with the analysis of cocitations. The coupled use of both cartographic procedures results in an enhancement of the visualization power.

**SPECULAR SYMMETRY:** The graph that the software produces is the following one (for a citation threshold GCS >= 20.)

**Graph 3: Barbacid 1981-1986. Year-by-year historiograph.**

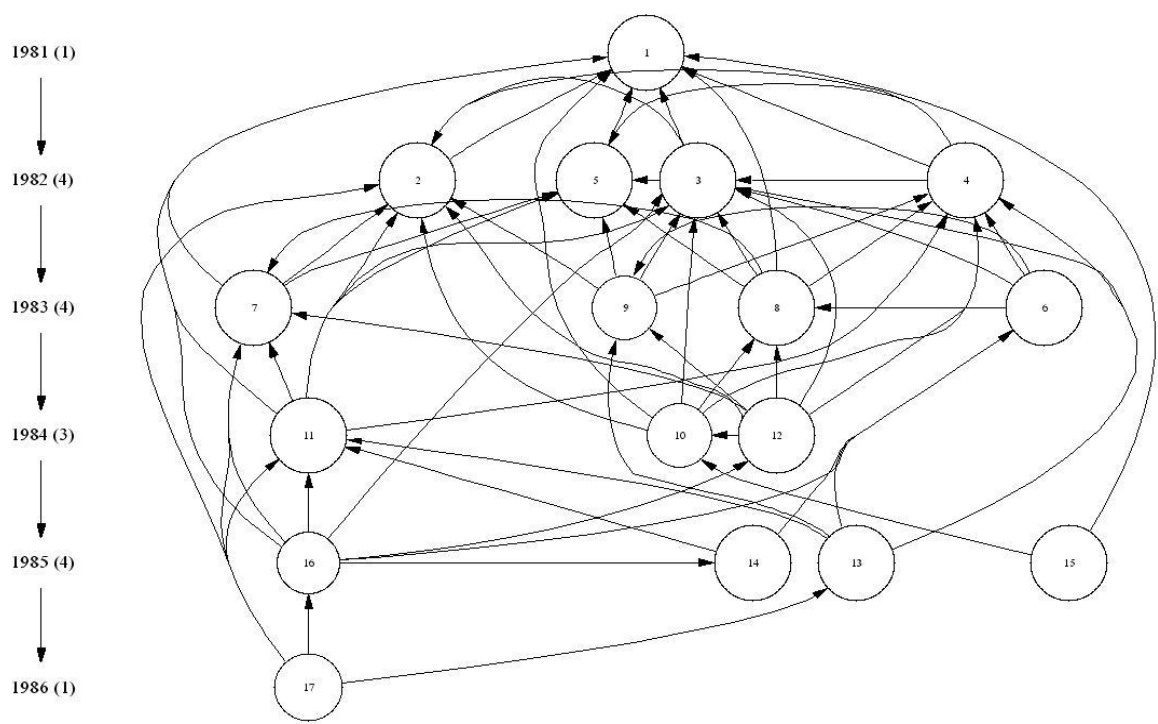

Because it is a year-by-year graph, the historio-bibliography is not visually displayed after an axis of symmetry. An axis that is not repeated in the other side of the symmetry is present in the maps of cocitations because the distances are constructed relatively to the thematic centrality of the co-cited articles. The dychotomous logic of the temporal relationship (an article is published in a year and not in any other), favors the speculation in the dual terms of symmetry dichotomy.

In the Table, the citation frequencies of the 5 main articles are presented. Its citation network clusters hierarchically the crucial events of the program that are studied in the 'Llull' paper. They are outstanding because they have been included by ISI as a part of the research fronts with codes 83-1740 ("Oncogenes and the genetics of human; viral transforming genes and their DNA structure") and 84-4046 ("Characterization of human and murine cellular oncogenes"). This is an specific condition (to his own history [30]) of production that, inside the regular flow of arrows in the graph, favors the main path analysis.

**Table V: Citation frequency of the articles in Graph II coded by ISI inside the research fronts.**

| Documento | Nº | Código ISI | Frecuencia de cita |
|---|---|---|---|
| 1983 NATURE 306(5944):658-661 SUKUMAR S; NOTARIO V; MARTINZANCA D; BARBACID M *INDUCTION OF MAMMARY CARCINOMAS IN RATS BY NITROSO-METHYLUREA INVOLVES MALIGNANT ACTIVATION OF H-RAS-1 LOCUS BY SINGLE POINT MUTATIONS* | 6 | 84-4046 | 654 |
| 1982 NATURE 296(5856):404-409 GOLDFARB M; SHIMIZU K; PERUCHO M; WIGLER M *ISOLATION AND PRELIMINARY CHARACTERIZATION OF A HUMAN TRANSFORMING GENE FROM T24 BLADDER-CARCINOMA CELLS* | 2 | 83-1740 | 441 |
| 1983 PROCEEDINGS OF THE NATIONAL ACADEMY OF SCIENCES OF THE UNITED STATES OF AMERICA-BIOLOGICAL SCIENCES 80(8):2112-2116 SHIMIZU K; GOLDFARB M; SUARD Y; PERUCHO M; LI Y; KAMATA T; FERAMISCO J; STAVNEZER E; FOGH J; WIGLER MH *3 HUMAN TRANSFORMING GENES ARE RELATED TO THE VIRAL RAS ONCOGENES* | 8 | 84-4046 | 278 |
| 1984 SCIENCE 223(4637):661-664 SANTOS E; MARTINZANCA D; REDDY EP; PIEROTTI MA; DELLAPORTA G; BARBACID M *MALIGNANT ACTIVATION OF A K-RAS ONCOGENE IN LUNG-CARCINOMA BUT NOT IN NORMAL TISSUE OF THE SAME PATIENT* | 12 | 83-1740 | 249 |
| 1982 PROCEEDINGS OF THE NATIONAL ACADEMY OF SCIENCES OF THE UNITED STATES OF AMERICA-BIOLOGICAL SCIENCES 79(9):2845-2849 PULCIANI S; SANTOS E; LAUVER AV; LONG LK; ROBBINS KC; BARBACID M *ONCOGENES IN HUMAN-TUMOR CELL-LINES - MOLECULAR-CLONING OF A TRANSFORMING GENE FROM HUMAN BLADDER-CARCINOMA CELLS* | 5 | 83-1740 | 237 |

**INTERNAL POLARITIES:** Transient and continuant indicators [31] inside the research front, are defined features of the evolution models in scientometrics [32]. One of its applicability domains is the professional mobility of the scientists. It is not a property that evaluates the absolute size (in terms of citations and co-citations) of the nodes, but of their homogeneity along the time (their duration as research publications). Inside the demographic pyramid those authors that are "alive" (they publish and they are cited) along the time period of analysis are the core authors, those that are "dying" along the period will be the continuant authors, and those only launched for a year are the "transient authors".

HistCite[TM] works on the basis of the model of circles. The area of each circle is proportional to the number of articles that cites to the one pointed out with a number inside the circle. With the help of a list of authors, produced by HistCite[TM] "Ranked all-author list" (see Table VI) and ordered by LCS, we have a criterion of local citation to the collection.

**Table VI: A list of the 10 first authors that are involved in the program "1988 Josef Steiner Award" that was described by the paper in Llull [29]. The authors are ranked by the local citation frequency towards the studied bibliographic collection.**

| # | Name | TLCS | TLCS/t | TGCS | TGCS/t | TLCSb | TLCSe | Pubs | TLCR |
|---|------|------|--------|------|--------|-------|-------|------|------|
| 1 | BARBACID M | 31 | 7.23 | 3810 | 1128.35 | 3 | 10 | 8 | 30 |
| 2 | PERUCHO M | 28 | 5.93 | 1865 | 502.37 | 6 | 6 | 6 | 16 |
| 3 | SANTOS E | 27 | 5.98 | 2364 | 568.85 | 3 | 6 | 6 | 24 |
| 4 | GOLDFARB M | 26 | 5.27 | 1407 | 292.87 | 6 | 5 | 4 | 9 |
| 5 | SHIMIZU K | 26 | 5.27 | 1407 | 292.87 | 6 | 5 | 4 | 9 |
| 6 | WIGLER M | 23 | 4.52 | 1129 | 223.37 | 6 | 5 | 3 | 4 |
| 7 | PULCIANI S | 19 | 4.25 | 1052 | 273.25 | 3 | 4 | 4 | 13 |
| 8 | FOGH J | 13 | 2.42 | 720 | 143.17 | 4 | 2 | 2 | 5 |
| 9 | REDDY EP | 11 | 2.48 | 1435 | 326.35 | 0 | 3 | 3 | 15 |
| 10 | LAMA C | 10 | 1.67 | 442 | 73.67 | 4 | 2 | 1 | 0 |

On the total of 46 participant authors, the identified demographic groups in the 'Llull' paper [29, p.542] are in the positions (see Table VI) N°1 and N°2 for the core authors; N° 3, 17, 18 and 19 for the continuant authors N° 20, 27, 37 and 41, 43 for the transient authors.

**The Kendall Question**

The central limit theorem permits the estimation of a probable magnitude on the difference between the sample mean and the mean of the population, and to fix the necessary size of the sample to obtain reliable estimations. The sample mean of a population under the conditions of the central limit theorem is normally distributed. Nevertheless, the probability distributions specific to the information processes do not satisfy these conditions (and, they frequently adjusts the hyperbolic distributions). To clarify the connection of these distributions M.G. Kendall, a student of the random number sampling (1940), made the question (1960): if he were to extend the publication period on a fixed subject from one year to two years, can the additional journals be estimated? [33] B.C. Brookes answered after 15 years (1975) by posing a sample theorem for the discrete finite distribution (based on the Taylor theorem of the basic mathematical analysis). [34]

To inquire the performance of the HistCite$^{TM}$ software when facing sampling problems, it was decided to experiment the temporal congruency in a duplication process. We formulate the Kendall question after the same subject than B.C. Brookes in 1975, ie: if we have a bibliographic data set on the "muscle fiber" in a year, is it possible to estimate the number of additional journals that probably contribute on the subject in two years?

The sampled values (n° of journals) were produced in response to a WOK search (with date 11/10/2004) for TS = Muscle Fiber (where TS = Topic), for an annual bibliography 2002. And the number of journals concerning the combined bibliography in two years (2002 + 2003) is predicted. When using the table "Ranked Source List", produced by HistCite$^{TM}$ after the specific indicators of local and global total citations scores (TLCS and TGCS) and after the frequency of publication, we ranked the sample by the frequencies. The total number of journals that have contributed to the "Muscle Fiber" specific subject in the year 2002 is $R(A)_{2002} = 152$; and in the year 2003 is $R(B)_{2003} = 133$ (see Table VII).

**Table VII: Journals that publish on "muscle fiber" in the year 2002 (bibliography A) and in 2003 (bibliography B), displayed by their frequencies of publication.**

| R | 1 | 2 | 3 | 4 | 5 | 6 | 7 | 8 | 9 | 10 | 11 | 12 | 13 | 14 | 15 | 16 | 17 | Total |
|---|---|---|---|---|---|---|---|---|---|----|----|----|----|----|----|----|----|-------|
| fr(A)$_{2002}$ | 114 | 23 | 7 | 2 | 2 | | | 1 | | | 2 | 1 | | | | | | 152 |
| fr(B)$_{2003}$ | 96 | 17 | 8 | 5 | 1 | 1 | 1 | | 3 | | | | | | | | 1 | 133 |

HistCite$^{TM}$ provides an important tool to merge bibliographies, because it admits specific procedures linked to the interdisciplinarity and the time period considered. And with the option 'Add Set', the journals from both bibliographies are cumulated effectively (see Table VIII).

**Table VIII: HistCite$^{TM}$ produces a list of journals with possibilities of arrangement according to bibliometrics criteria (local and global citations, frequency of publication) – Extract of the Top 10 journals, for the bibliography that results from the merging of A and B.**

| # | Name | TLCS | TGCS | Pubs |
|---|------|------|------|------|
| 1 | JOURNAL OF APPLIED PHYSIOLOGY | 10 | 105 | 28 |
| 2 | MUSCLE & NERVE | 4 | 83 | 21 |
| 3 | MEDICINE AND SCIENCE IN SPORTS AND EXERCISE | 8 | 80 | 20 |
| 4 | FASEB JOURNAL | 0 | 1 | 15 |
| 5 | JOURNAL OF ELECTROMYOGRAPHY AND KINESIOLOGY | 7 | 56 | 12 |
| 6 | JOURNAL OF ANIMAL SCIENCE | 1 | 17 | 10 |
| 7 | AMERICAN JOURNAL OF PHYSIOLOGY-CELL PHYSIOLOGY | 0 | 15 | 8 |
| 8 | JOURNAL OF BIOLOGICAL CHEMISTRY | 2 | 72 | 8 |
| 9 | AMERICAN JOURNAL OF PHYSIOLOGY-ENDOCRINOLOGY AND METABOLISM | 0 | 41 | 7 |
| 10 | EUROPEAN JOURNAL OF APPLIED PHYSIOLOGY | 0 | 10 | 6 |

The combined bibliography (A + B), an approach of the publications on "muscle fiber" included in WOK for the years 2002 and 2003, listed a total of $R(A+B)_{2002+2003} = 231$ journals.

By appliance of the sample theorem proposed by Brookes to solve the Kendall Question [34], the estimation of the additional number of journals, M, prognosticated for a search period extended from 1 year to 2 years, should result from the calculation (after the data provided by $fr(A)_{2002}$ in the Table VII and by sum of the alternate positive and negative values):

$$M(A)_{2002} = 114 - 23 + 7 - 2 + 2 - 1 + 2 - 1 = 98$$

The predictor value of $M(A)_{2002}$ was not enough good because:

$$R(A)_{2002} + M(A)_{2002} = 152 + 98 = 250 <> 231 = R_{2002+2003}(A+B).$$

Thus simple errors are detected, like Brookes indicates in the page 28 of the article. [34]

To estimate the sampling errors we use the 'Outer References' tables, exterior nodes to the original bibliography automatically calculated (see Table IV for an employ of the table 'Missing Links'). The method will be the elaboration of a new bibliography for 2002 + 2003 that makes use of a number of journals more close to the estimated '250'.

In this case the tables of exterior nodes, for bibliography $(A)_{2002}$ and for bibliography $(B)_{2003}$, produce 8007 and 8024 references respectively (see in Table IX an extract of the 'Outer References' for the 2002 bibliography).

**Table IX: 'Outer References' – Top ten nodes exterior to the original bibliography (A)$_{2002}$ (over a total of 8007)**

Glossary   HistCite Guide

ISI Web of Science location: |_______________|
Cited references outside of this network.

Total: 8007 (top | 8007 | shown).   [Re-display]

Sorted by **LCS**.

| # | LCS | Reference |
|---|-----|-----------|
| 1 | 10 | BROOKE MH, 1970, ARCH NEUROL-CHICAGO, V23, P369 WoS |
| 2 | 9 | BARSTOW TJ, 1996, J APPL PHYSIOL, V81, P1642 WoS |
| 3 | 9 | PETTE D, 1997, INT REV CYTOL, V170, P143 WoS |
| 4 | 8 | CHIN ER, 1998, GENE DEV, V12, P2499 WoS |
| 5 | 8 | SCHIAFFINO S, 1989, J MUSCLE RES CELL M, V10, P197 WoS |
| 6 | 7 | BARSTOW TJ, 2000, EXP PHYSIOL, V85, P109 WoS |
| 7 | 7 | BROOKE MH, 1970, J HISTOCHEM CYTOCHEM, V18, P670 WoS |
| 8 | 7 | WU H, 2000, EMBO J, V19, P1963 WoS |

| 9 | 7 | `SCHIAFFINO S, 1996, PHYSIOL REV, V76, P371` WoS |
| 10 | 6 | `ARMSTRONG RB, 1984, AM J ANAT, V171, P259` WoS |

For each table, the only selected entries were those with year of publication 2002 in A or 2003 in B. 96 additional references were considered for the year 2002, and 117 for 2003. As the items appear ranked by bibliometrics criteria (after its local citation frequency, LCS), the rule 80/20 [35] can be applied. And it establishes that 20% of the more important sources contain 80% of the total items. Thus, the new references to be included in the bibliography $(A)_{2002}$ can be limited to 19 and, to 23 the references to be added to the information production process for 2003, $(B)_{2003}$.

By using the option 'Add' from the program interface screen 'Add set', all the data are collected, by adding the additional records for the two years. The final number of journals obtained for the merged bibliography was R2002+2003(A+B) = 243. The difference with the calculated amount coming from the solution by Brookes R(A)2002 + M(A)2002 = 250, has been reduced from 7.6% to 2.8%.

## CONCLUSIONS

HistCite[TM] is an instrument for the co-citations analysis between authors, singularly adequate to the research of the intellectual structure and the history. Based in the technic of the reference analysis by M.M. Kessler and in the co-citation analysis introduced by H. Small in 1974, it is the last result of the algorithmic historiography that was evolved from the computational linguistic [36] by E. Garfield in 1963 in the framework of the school of Derek John de Solla Price [37].

HistCite[TM] offers the elaboration of histograms, or histories of citation based on nodes diagrams and connections, and into schemes of dissemination. This capacity works with citation matrices (an input-output resource also referred in diverse occasions by Price [38][39]). By enhancing the expression possibilities of the critical path method, HistCite[TM] deepens the reflection by Garfield about the citation cycle by Price [40].

HistCite[TM] decisive relevance for the evaluation comes from the fact that its power of visualization is reinforced by the employ of indicators. HistCite[TM] introduces 8 bibliometrics indicators. Its combined use adds more dimensions to the information that it explores. With a workload capacity of 500 references, although with no limitations in the successive implementation of consecutive mobile windows with the same size, HistCite[TM] works in real time. The system HistCite[TM] is presented here as an application of the bibliographic coupling methodology and the cocitations analysis that evaluates, for a fixed temporal interval, the citation networks internal to a historio-bibliographic collection.

# INFORMATIONAL ABSTRACT

## Aims

the presentation of the advantages of the new bibliometrics configuration option of HistCite[TM] (2004) for the identification of papers;

to show the additivity of the reference levels according to the reading domains of an author;

to determine where the number of articles generated per author is higher than per coauthor;

to show how when pointing out a citation year for an author the selection of a previous (or posterior) year (s) can be an advantage for him by means of the examination of two graphs yielded by the software;

to model the feasibility of a paper as a reference, starting from the citations that it receives;

considering that the obsolescence gives rise to an effect of cumulative advantage, it is approached the extraction of a graph that explicitly shows this effect in terms of citations to old frequently cited articles;

to show which way inside the graph maker option the change in the initial step results in that the papers can not only be grouped under subject but also with a time-based criteria;

to approach the identification of the nuclear, continuant and transient scientific authors community by using the HistCite[TM] Graphmaker option;

to discuss the software difficulties at the time of giving graphical sense.

## Methodology

Exploitation of the new indicators for the global and the local citation score as a function of time (GCS/t, LCS/t), facilitated by the 2004 version of HistCite[TM], starting from the same initial bibliographic set. Examination of the central limit theorem appliance conditions based on the graphic modelization of the Kendall question.

## Results

To study the history of the journal 'Management Science' Editorial Department an histogram was elaborated. It expressed the local citation score in the end of the period (LCS$_e$) 1970-2005 at the abcisas axis and the local citation score in the beginning of the same period (LCS$_b$) at the axis of the ordenates. Some of the relevant key papers were observed , what featured some 50% of the selected candidacies in the case study paper published by the scientific press. Both couple of indicators selected the same material, by assigning main relevance to the same initial author. This identified person is the same than the original leader as reported by the published article. Perhaps the second pair of indicators (GCS/t, LCS/t) is more informative because it scales all the other authors by assigning a discrimination criterion based on the subject. The theoretical or methodologic articles were only selected, and no empirical material nor derived from behaviour was obtained.

The comparative study of the graphs produced by the software was induced by the employ of HistCite[TM] to map the 1988 Dr Josef Steiner Cancer Research Foundation Award, investigating the period 1981-1986. The logical content of the bibliography was extracted through two ISI research fronts: code 83-1740, Oncogenes and the genetics of human; viral transforming genes and their DNA structure; and code 84-4046, Characterization of human and murine cellular oncogenes. Although the case history has been published, was expressed in terms of professional anticipacionism and used the Weibull distribution, HistCite[TM] generates a remarkably similar structure based on co-citations. The crucial events, identified in terms of acyclic graphs by HistCite[TM] also are the same than those underlined by a second publication that determines the connectivity into citation networks based in a study of similar sources. The automatic algorithmic model correctly analyzes the implied scientific authors demography. And is useful when identifying the research core population of authors, the research continuants along the whole process, and the population of those transients which assist in the elaboration of the results.

When inquiring on the graphic sensibility of the HistCite[TM] software graphic performance a visual answer to the Kendall question was tried (if we have a bibliographical data set for a year, the number of additional journals probably contributing to this set for the next two years can be estimated?). An annual bibliography was chosen – 2002 – and the number of journals for the combined bibliography 2002+2003 was pronosticated. Like in the original paper that poses this question for finite discrete distributions, the selected subject was the "muscle fiber". After the 'Ranked Source List' table from HistCite[TM], the 2002 journals were ranked by frequencies, and a total of 152 was signalled. By application of the formulation for the amount of additional journals, M, under the Central Limit Theorem assumptions, it was obtained that $M(2002) = 98$. In fact,the combined bibliography (2002 + 2003) – easily produced with the HistCite[TM] 'Add Set' option from the WOS records – listed a total of 231 journals. The value of the prediction, $M(2002)$, is only partially correct because $152 + 98 = 252$. The graphs that were produced for the bibliographies 2002 vs. 2002+2003 shows that : several selected records do not satisfy the annual employed criterion. Thus the graphs were employed to refine the bibliographies. Once the journals not belonging to the year were identified, the references were manually deleted. The 'Graph Maker' option seems not prepared to manage with a selection of nodes (marked) from the main table. What makes it difficult to reformulate the question once posed. When answering the Kendall question, for instance, although the month of publication is facilitated in the articles of the bibliography, a graph is not immediately available for the papers of the first semester as something opposite to the graph that could be made with the published articles along the a second semester; what forbids the Kendall question formulation in its original terms.